# Measurements and scaling of X-ray total scattering from single crystals

S. Gorfman[*,1], M. Eremenko[2,3], V. Krayzman[2], A. Bosak[4], P. Y. Zavalij[5], I. Levin[2*]

[1]Department of Materials Science and Engineering, Tel Aviv University, Tel Aviv, Israel

[2]Materials Measurement Science Division, National Institute of Standards and Technology, NIST, Gaithersburg MD 20899 USA

[3]Spallation Neutron Source, Oak Ridge National Laboratory, Oak Ridge, TN USA

[4]European Synchrotron Radiation Facility, Grenoble, France

[5]Department of Chemistry and Biochemistry, University of Maryland, College Park MD 20742 USA

*Corresponding authors: gorfman@tau.ac.il; igor.levin@nist.gov

**Abstract**

We present a measurement protocol and a data reduction workflow for obtaining single-crystal X-ray total-scattering datasets that capture both Bragg-peak and diffuse scattering intensities on an absolute (electrons$^2$/atom) scale. We further demonstrate that the intensity scale factor derived from crystallographic refinements using Bragg peaks is in close agreement with that obtained by matching the scattering function, computed via spherical integration of the 3D total scattering signal, to the theoretical coherent baseline, calculated as the spherical average of the Debye-Waller factor. The latter approach can be applied to diffuse scattering without including Bragg peaks. Our results set the ground for structural refinements using large atomic configurations while simultaneously fitting Bragg intensities and diffuse scattering from a single crystal. Moreover, with the convergence between the two scaling methods, the Bragg and diffuse components can be obtained from the same total-scattering dataset, as achieved in this work, or measured independently.

ORCID:

Igor Levin: 0000-0002-7218-3526

Maksim Eremenko: 0000-0002-2875-968X

Victor Krayzman: 0000-0001-9010-2681

Semën Gorfman: 0000-0003-1652-9788

# 1. Introduction

X-ray or neutron total scattering performed on powder samples has become a common method for analyzing the local and nanoscale structure in crystalline materials. The total scattering includes Bragg peaks that represent the average, infinitely periodic atomic arrangement, along with a smoothly varying diffuse background, which contains information about locally correlated deviations from this average (Egami & Billinge, 2003). The Fourier transform of the total signal produces an atomic pair distribution function (PDF). Since interatomic correlations are often anisotropic, the resulting diffuse scattering is also anisotropic (Welberry, 2004). Orientational averaging inherent in powder measurements masks this anisotropy, limiting the amount of information that can be extracted from such data.

Recent results demonstrated that combining one-dimensional (1D) total scattering data from powders with a three-dimensional (3D) diffuse scattering from a single crystal permits recovering occupational and displacement correlations that would be undetectable from powder data alone (Eremenko *et al.*, 2019; Krayzman *et al.*, 2022; Eremenko *et al.*, 2025). Such combined analysis, while powerful, depends on having both forms of the same material, assumes identical elemental composition and structure for both samples, and requires coordinating measurements at different facilities and instruments, each demanding specialized expertise. Measurements under nonambient conditions add complications, as variations in sample sizes and environments can affect controlled parameters. Therefore, if single crystals are available, it would be advantageous to have the ability to generate quantitative models of correlated atomic disorder across multiple length scales without needing powder measurements.

One emerging approach toward this goal is a 3D $\Delta$PDF (Weber & Simonov, 2012), which represents the Fourier transform of the diffuse intensity with Bragg peaks removed. A 3D $\Delta$PDF exhibits positive and negative peaks at or around locations that correspond to distance vectors connecting different pairs of atoms in the average structure. These peaks reflect deviations from the average probability of finding a specific pair. Analyzing signs, intensities, locations, and shapes of these peaks allows for the determination of correlations underlying diffuse scattering. While powerful, existing analysis methods for 3D $\Delta$PDF do not produce fully atomistic structural models that would show explicitly how the local structure evolves into the average.

Another solution could involve measuring both Bragg peaks and diffuse scattering from a single crystal in one dataset and using it to refine large atomic configurations, similar to the current use of powder total scattering data with the Reverse Monte Carlo (RMC) minimization method (Tucker *et al.*, 2007; Eremenko *et al.*, 2017; Zhang *et al.*, 2020). Such data, potentially augmented by direction-resolved X-ray absorption fine structure (Krayzman *et al.*, 2009), can be expected to yield highly detailed structural models. The key experimental hurdle is achieving an adequate signal-to-noise ratio (SNR) for the weak diffuse component while avoiding detector saturation by intense Bragg peaks – a problem that has not been adequately addressed so far. Koch et al. (Koch *et al.*, 2021) considered various artefacts affecting single-crystal total scattering measurements and developed data-processing protocols to mitigate them. Their focus, however, was on producing an artefact-free diffuse scattering component, without addressing the accuracy of Bragg peak intensities. Moreover, some artefacts, such as blooming, which are relevant for the amorphous silicon area detector used in their study, become largely insignificant for the single-photon-counting hybrid pixel detectors increasingly adopted at synchrotron facilities.

In essence, combining integrated Bragg peak intensities and 3D diffuse scattering from a single crystal in structural refinements does not require having both components in the same dataset. From the practical standpoint, it would be computationally more effective to introduce them separately, with Bragg peaks included out to large momentum transfers to provide sensitivity to atomic probability density distributions, and diffuse intensities over more limited reciprocal-space volumes sufficient to uncover correlations but amenable to fitting using current computing power.

Regardless of the analysis approach, quantitative fitting of diffuse scattering, whether in reciprocal space or after the Fourier transform, requires placing intensities on an absolute scale (e.g., electrons$^2$/atom). When Bragg peaks are acquired together with the diffuse signal, they can be used to calibrate the scale by performing crystallographic refinements of the average structure. Alternatively, as has been shown recently (Eremenko *et al.*, 2025), one can exploit the asymptotic behavior of the scattering function obtained from the spherically integrated diffuse scattering to set the scale of diffuse scattering without Bragg intensities.

Here, we describe a workflow for obtaining single-crystal X-ray total scattering data using a synchrotron beamline optimized for the detection of weak diffuse scattering. We demonstrate an approach for dealing with detector saturation by Bragg peaks and for subtracting the Compton scattering, which contributes significantly to the scattering intensity at large momentum transfers. We further show the convergence of the two methods for scaling the intensity of diffuse scattering outlined above. Our validation of the approach relying on the known asymptote of the scattering function enables robust scaling of single-crystal diffuse intensities, opening the door to flexible combined analysis of Bragg and diffuse scattering even when these signals are collected using different instruments.

**2. Experimental**

**Test system**

In the absence of reference materials with well-established atomic probability density distributions and interatomic correlations, we selected a canonical relaxor ferroelectric $PbMg_{1/3}Nb_{2/3}O_3$ (PMN) (Bokov & Ye, 2006) as a test system because of the availability of extensive high-quality experimental data and results of large-box structural refinements from a combined fitting of these datasets (Eremenko *et al.*, 2019). PMN exhibits a relatively simple cubic perovskite average structure, but highly complex nanoscale correlations that generate prominent anisotropic diffuse scattering (Xu *et al.*, 2004). We used single crystals of PMN from our previous studies (Eremenko *et al.*, 2019, 2025).

**Synchrotron X-ray total scattering measurements**

X-ray scattering intensity from a PMN single crystal was measured at the diffraction side station of the ID28 beamline of the European Synchrotron Radiation Facility (Girard *et al.*, 2019). This station is equipped with a HUBER four-circle goniometer, a PILATUS[1] 1M pixel area detector, and the single-crystal X-ray diffractometry tools for crystal preparation and mounting. The measurements were conducted at ambient temperature using X-rays with a wavelength of 0.6968 Å. The approximately cylindrical crystal (≈ 30 μm in diameter) was glued to a glass needle and mounted onto the sample holder. Scattering intensity was recorded in the "continuous" mode, in which the crystal rotates at a constant angular speed around the diffractometer $\varphi$-axis, integrating the detector images over $\Delta\varphi = 0.125°$ rotation step and the exposure time $\tau = 0.25$ s. Full 360° rotations were repeated at four different detector angles, achieving a maximum scattering angle $2\theta_{max} = 107°$ and momentum transfer $Q_{max} \approx 14$ Å$^{-1}$. This measurement was repeated with six different primary beam absorbers spanning five orders of magnitude in beam attenuation. The degree of attenuation of the primary beam was determined for each absorber using a beam monitor. The experiment yields $J_{\text{lab}}(x_d, y_d, \varphi)$ representing the number of photons accumulated inside the detector pixel $x_d y_d$ during the crystal's rotation over the angular range $[\varphi, \varphi + \Delta\varphi]$ within the time interval $\tau$ and using the primary beam absorber, corresponding to the beam monitor value $B$.

We used CrysAlisPro (Meyer & IUCr, 2015) for the initial inspection of the scattering intensity and to determine the orientation matrix. The rest of the data analysis was performed using custom MATLAB scripts that provided the necessary flexibility and self-consistency. These scripts are supplied as supplementary information. The first stage of data processing involved determining centers of mass for the Bragg reflections and refining the orientation matrices for datasets acquired using different beam absorbers. During this refinement, a cubic lattice constraint was applied, allowing for the adjustment of a single lattice parameter along with three Euler angles. Subsequently, we performed the reciprocal space reconstruction (RSR) as described below.

**Laboratory X-ray diffraction measurements**

A reference set of Bragg peak intensities was collected with a Bruker D8 Venture diffractometer, where a small fragment of the PMN crystal, about 0.14 × 0.08 × 0.025 mm$^3$, was measured at ambient temperature (298 ± 2 K) using Mo K$\alpha$ radiation (0.71073 Å). The integrated intensities were obtained in the Bruker Apex5 software and corrected for absorption using SADABS. The minimum and maximum transmission values were 0.121 and 0.266, respectively. The measurements covered the $hkl$ range of $-8 \leq h, k, l \leq 8$ out to $Q_{max} = 12.9$ Å$^{-1}$, yielding a total of 7463 reflections, of which 90 unique reflections ($R_{int} = 0.032$, $R_{sig} = 0.0154$) were used in the structural analysis.

## 3. Methodology for processing 3D total scattering data

Figure 1 presents a workflow transforming raw intensities into absolute-scale data for quantitative analysis. The subsequent sections detail each step.

---

[1] Certain equipment, instruments, software, or materials, commercial or noncommercial, are identified in this paper in order to specify the experimental procedure adequately. Such identification is not intended to imply recommendation or endorsement of any product or service by NIST, nor is it intended to imply that the materials or equipment identified are necessarily the best available for the purpose.

| | | | |
|---|---|---|---|
| 1 | Raw data → $J_{\text{lab}}(x_d, y_d, \phi)$ | 8 | Compton and baseline → $I_{\text{compt}}$, $S_{\text{calc}}^{\text{base}}(Q)$ |
| 2 | Data corrections → $J'_{\text{lab}}(x_d, y_d, \phi)$ | 9 | Determining scale factors → $s, p$ |
| 3 | Reciprocal space reconstruction → $I_{\text{cryst}}(H, K, L)$ | 10 | Placing on the absolute scale → $I_{\text{cryst}}^s(H, K, L)$ |
| 4 | Calculating uncertainties → $\sigma(H, K, L)$ | 11 | Fitting background around Bragg → $I_0(H, K, L)$ |
| 5 | Replacing saturated voxels → $I/\sigma < 10^3$ | 12 | Integrating Bragg intensities → $I_{hkl}^{\text{obs}}$ |
| 6 | Visual inspection (layers/rods) | 13 | Selecting absorbers for Bragg peaks |
| 7 | Spherical averaging → $I(Q)$ | 14 | Structural refinement |

**Figure 1.** Workflow for converting raw X-ray scattering intensities into absolute-scale data. The intensity notations correspond to those used in the equations presented in the subsequent sections.

### 3.1 Reciprocal space reconstruction

The first objective of RSR is to transform the instrument-related coordinates $x_d, y_d, \varphi$ into the components of the scattering vector $\boldsymbol{Q}$ (Figure 2) and to convert the measured scattering intensity $J_{\text{lab}}(x_d, y_d, \varphi)$ into the $I(Q_x, Q_y, Q_z) = I_{\text{cryst}}(H, K, L)$ look-up tables. Here, $Q_x, Q_y, Q_z$ / $H, K, L$ denote the components of the scattering vector in either the Cartesian laboratory $\boldsymbol{e}_i$ or reciprocal crystallographic $\boldsymbol{a}_i^*$ coordinate systems (see Appendix A and Figure 2 for more details) and at zero diffractometer angles:

$$\boldsymbol{Q} = Q_x \boldsymbol{e}_1 + Q_y \boldsymbol{e}_2 + Q_z \boldsymbol{e}_3 = 2\pi(H\boldsymbol{a}_1^* + K\boldsymbol{a}_2^* + L\boldsymbol{a}_3^*) \qquad (1)$$

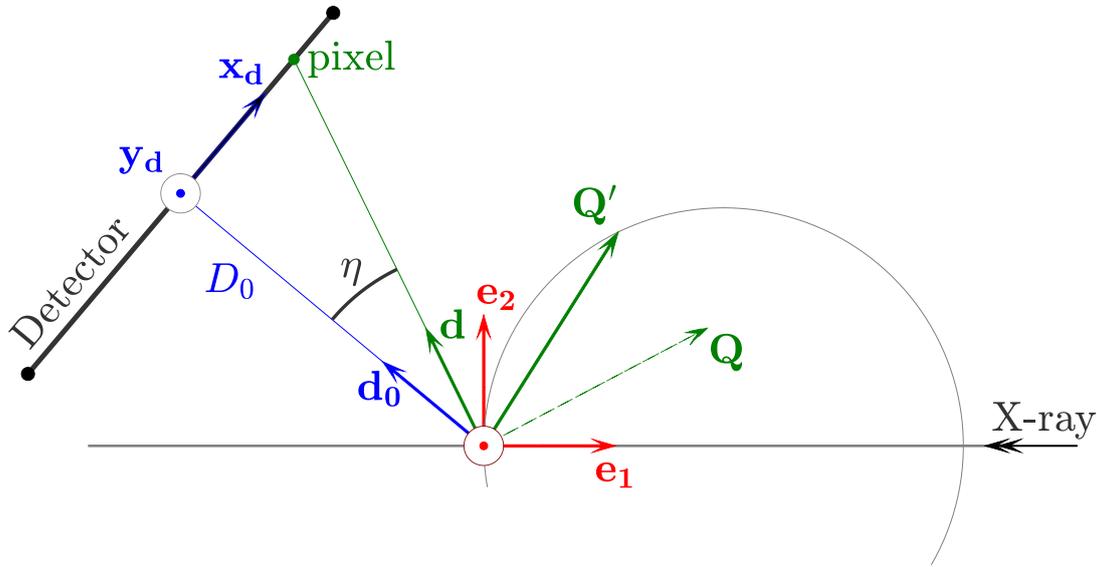

**Figure 2.** Schematic illustration of the diffraction experiment, introducing notations used in the text. The image shows the primary X-ray beam, Cartesian laboratory coordinate system $\boldsymbol{e}_1$ $\boldsymbol{e}_2$ ($\boldsymbol{e}_3$ is defined as their cross product). $\boldsymbol{d}_0$ and $\boldsymbol{d}$ are the unit vectors defining the direction of the detector normal and the position of a given detector pixel, respectively. The two vectors are separated by the oblique incidence angle $\eta$. $D_0$ is the distance to the detector plane. $\boldsymbol{x}_d, \boldsymbol{y}_d$ are the unit vectors, defining the detector axes. $\boldsymbol{Q}$ and $\boldsymbol{Q}'$ are the scattering vectors before and after rotation.

The second objective is to correct $J_{\text{lab}}(x_d, y_d, \varphi)$, enabling its conversion to the coherent scattering intensity $I_{\text{coh}}(\mathbf{Q})$, which can be expressed as

$$I_{\text{coh}}(\mathbf{Q}) = \frac{1}{N} \sum_m \sum_n f_m f_n^* \exp\{i\mathbf{Q}\mathbf{R}_{mn}\}. \tag{2}$$

Here $N$ is the total number of atoms in the irradiated volume of the crystal, $f_m = f_m(Q)$ are atomic scattering factors, and $\mathbf{R}_{mn} = \mathbf{R}_m - \mathbf{R}_n$ is the vector connecting atoms $n$ and $m$.

Our RSR implements the necessary data correction and normalization steps according to the kinematical theory of X-ray scattering (Warren, 1990; Guinier, 1994). The first step applies pixel-by-pixel (polarization, Lorentz, and oblique incidence) and flat-field (exposure time $\tau$, primary beam monitor $B$ and the distance to the detector $D_0$) corrections:

$$J'_{\text{lab}}(x_d, y_d, \varphi) = J_{\text{lab}}(x_d, y_d, \varphi) \, R(x_d, y_d, \varphi) \tag{3}$$

The correction factor $R$ is introduced as (see Appendix B for more details)

$$R(x_d, y_d, \varphi) = \frac{R_0 D_0^2}{B\tau \sin^2 \psi \cos^3 \eta} \, \delta V^* \tag{4}$$

Here $\psi = \psi(x_d, y_d)$ is the angle between the polarization direction of the primary beam and the propagation direction of the scattered beam, $\eta = \eta(x_d, y_d)$ is the oblique incidence angle (Figure 2). $R_0$ accumulates all the factors that remain constant throughout the experiment yet contribute to the scattered intensity. $\delta V^* = \delta V^*(x_d, y_d, \varphi)$ is the reciprocal space volume covered by the single 3D pixel, which can be calculated from the known functional dependence $Q_{x,y,z}(x, y, \varphi)$ (see Appendix A for more details):

$$\delta V^* = \left|\frac{\partial(Q_x, Q_y, Q_z)}{\partial(x, y, \varphi)}\right| S \, \Delta\varphi \tag{5}$$

where $S$ is the area of the detector pixel.

The next RSR step partitions reciprocal space into fixed-size voxels extending over the $\Delta H, \Delta K$ and $\Delta L$ reciprocal lattice units (r.l.u.) along $\mathbf{a}_1^*, \mathbf{a}_2^*, \mathbf{a}_3^*$, respectively. We assume that the voxel's volume is significantly greater than $\delta V^*$. The reconstruction algorithm converts $Q_x, Q_y, Q_z$ into $H, K, L$ and assigns all measured pixels to their corresponding voxels. At this stage, we also symmetrized the data using operations of the $m\bar{3}m$ point symmetry group, generating symmetric copies of each pixel in the $H, K, L$ space.

We then accumulate the volume-average voxel intensities $I_{\text{cryst}}(H, K, L)$ according to:

$$I_{\text{cryst}}(H, K, L) = \frac{\sum J'_{\text{lab}}(x_d, y_d, \varphi)}{\sum \delta V^*} \tag{6}$$

Here, both summations are over all pixels that fall into the $HKL$-based voxel. Appendix B shows that these reconstructed values are related to the scattering amplitude as:

$$I_{\text{cryst}}(H, K, L) = s\langle I_{\text{coh}}(H, K, L)\rangle \tag{7}$$

Here, $s$ is the scaling constant and the average is over the $HKL$-based voxel. Approaches to determining this constant will be considered in the next sections.

Our RSR procedure enables calculating the standard uncertainties $\sigma(H,K,L)$ of the $I_{\text{cryst}}(H,K,L)$ values, assuming the Poisson counting statistics so that $\sigma^2(x_d, y_d, \varphi) = J_{\text{lab}}(x_d, y_d, \varphi)$. Then, according to equation (6)

$$\sigma^2(H,K,L) = \frac{\sum J_{\text{lab}}(x_d, y_d, \varphi) R^2(x_d, y_d, \varphi)}{(\sum \delta V^*)^2} \qquad (8)$$

**3.2 The problem of detector oversaturation**

The simultaneous acquisition of both Bragg and diffuse scattering intensities may lead to saturation artefacts in certain pixels $J_{\text{lab}}(x_d, y_d, \varphi)$, particularly in the proximity of intense Bragg reflections. Moreover, any departure from the linear response of the detector leads to a systematic underestimation of recorded intensity values. The PILATUS active pixel area detector is known to maintain linearity at a photon count rate below $F_{max} = 10^6$ cps. This quantity is defined by the deadtime of the detector. Additionally, a hard cap is imposed on the total number of counts per pixel per detector frame, specified as $1.5 \times 10^6$ counts. This quantity is defined by the dynamic range of the detector. Consequently, a pixel should be considered as oversaturated if either of these limits – instantaneous count rate or total frame-integrated count – is exceeded.

In our experimental setup, each frame was acquired with an exposure time of up to 0.25 seconds. Thus, we conservatively define a threshold of $I_{max} = 10^5$ counts per pixel per frame, corresponding to SNR ≈ 300, as a safe upper limit to ensure unsaturated signal acquisition at the pixel level, provided that the intensity does not change much within a pixel itself. If the latter is true, then the intuitive approach to correcting the oversaturation is to replace the saturated pixels with equivalent ones from the data set measured using an adequate absorber. Unfortunately, this method is not feasible because even minor discrepancies in orientation matrices during successive measurements lead to different reciprocal space coordinates $HKL$ assigned to the same detector pixel. Therefore, we adopted a strategy that operates at the voxel level. Each voxel represents a volumetric element in reciprocal space that aggregates the intensity contribution of $10^3 - 10^4$ pixels. For voxel-based reconstructions, we define a higher saturation threshold corresponding to SNR≈1000, providing a robust criterion for identifying and mitigating saturation.

In this way, we produced a combined look-up table in which the slowly varying diffuse scattering component is free from saturation effects. Figures 3 and 4 present the results. Figure 3 displays the 2-dimensional reconstruction of $0KL$, $1KL$ and $KKL$ layers, each integrated over the range of 0.05 r.l.u. Figure 4 provides additional details of the scattering intensity, including its radial and angular dependence through the intensity profile along the cubic face diagonal direction, and the polar plot illustrating the anisotropy of the distribution.

This voxel-level criterion becomes unreliable near Bragg peaks, where photon fluxes are highly localized within exceedingly small angular volumes, often many orders of magnitude smaller than the extent of a single pixel in the reciprocal space. In such cases, the instantaneous count rate may exceed $F_{max}$ even if the total integrated intensity remains significantly below $I_{max}$. Thus, the detection of oversaturation based solely on the total intensity is ineffective, and the correction scheme described above is valid only for slowly varying diffuse scattering data. For Bragg peak intensities, we use a different approach as detailed in Section 3.4.

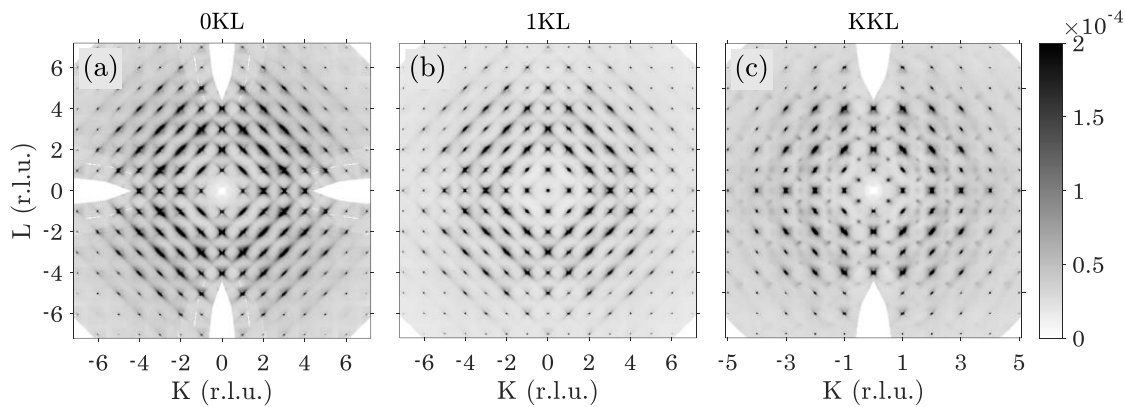

**Figure 3**. Grayscale maps of 2D reconstructions of X-ray scattering intensity in the $0KL$, $1KL$ and $KKL$ layers obtained from the combined (SNR < 1000) dataset, each integrated over the range of 0.05 r.l.u. The maps reveal both sharp Bragg peaks and strongly anisotropic diffuse scattering. Due to the extremely high and typically oversaturated intensity of the Bragg peaks, the grayscale was clipped at 0.0002 of the maximum intensity to enhance the visibility of the diffuse features.

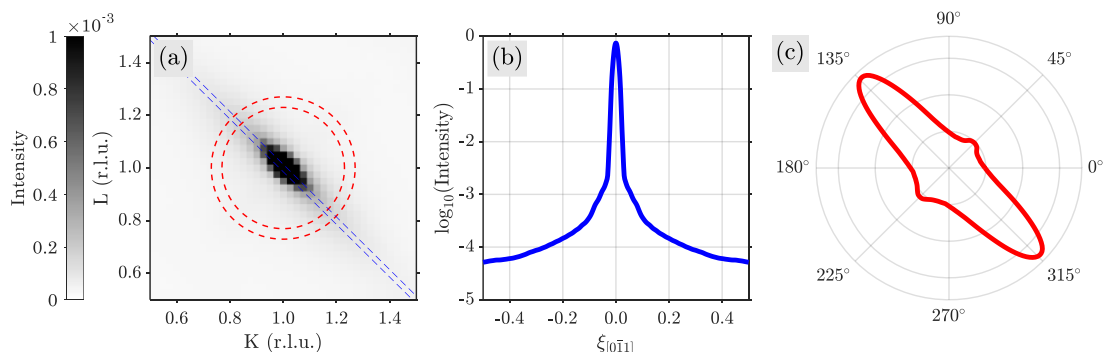

**Figure 4**. Details of the diffuse scattering intensity around the 011 Bragg reflection from the combined dataset (SNR < 1000 criterion). The figure is composed of three panels presenting complementary aspects of the reconstructed intensity distribution in the $0KL$ reciprocal lattice plane. (a) Reconstructed 2D intensity map around the Bragg peak, displayed over ± 0.5 r.l.u. A dashed red circular region marks the radial zone used for angular averaging, and dashed blue lines indicate the strip used to extract a one-dimensional profile along the diagonal direction. (b) Logarithmic intensity profile along the $[0\bar{K}K]^*$ reciprocal direction, averaged over a finite-width strip perpendicular to the path. (c) Polar plot of the angular dependence of the diffuse intensity, extracted from the region shown in a.

### 3.3 Placing data on the absolute scale

In traditional crystallographic refinements, the scale factor $s$ is treated as an independent variable and obtained by minimizing the difference between $I_{hkl}^{obs}$ and $sI_{hkl}^{calc}$. In contrast, refinements based on total scattering data using large atomic configurations require experimental data to be on an absolute scale (electrons$^2$/atom) so that no scale factor is involved in the fit. Including this factor in refinement, albeit permitted by the existing software, would lead to significantly incorrect structural information as it starts to compensate for the strength of correlations, etc. If both Bragg and diffuse scattering are measured simultaneously, this issue can be resolved by first performing a conventional crystallographic refinement to determine $s$, and subsequently rescaling the entire total scattering dataset, including its diffuse component.

Alternatively, $s$ can be determined by exploiting a known asymptotic behavior of the powder scattering function, as suggested by (Eremenko *et al.*, 2025). In this approach, the 3D total scattering data is first spherically averaged and converted into the scattering function, which is then matched to a theoretical baseline calculated from the composition and atomic displacement parameters ($U_{ij}$). These parameters need only be known approximately and can be adopted from literature, if available, or estimated from crystallographic refinements of Bragg intensities, either from the same total scattering dataset or from a separate experiment. Here, we demonstrate that the two scaling methods converge, yielding similar scale factors. The procedure includes the following steps:

**Step 1:** Integrating the 3D scattering intensity distribution $I_{cryst}(H,K,L) \equiv I(\boldsymbol{Q})$ over all the possible crystal orientations to obtain a 1D intensity function $I(Q)$ that would be measured for an equivalent powder specimen:

$$I(Q) = \langle I(\boldsymbol{Q}) \rangle_{|\boldsymbol{Q}|=Q} \qquad (9)$$

In practice, this integration is performed by dividing the $Q$-range of interest into a finite number of bins and averaging all $I(\boldsymbol{Q})$ values corresponding to each $|\boldsymbol{Q}|$ within a given bin. Figure 5 displays the result of such averaging for datasets collected with five different beam absorbers. The graphs are vertically offset for clarity, but the zero level for each graph is indicated by a dashed line and a corresponding label on the right-hand side.

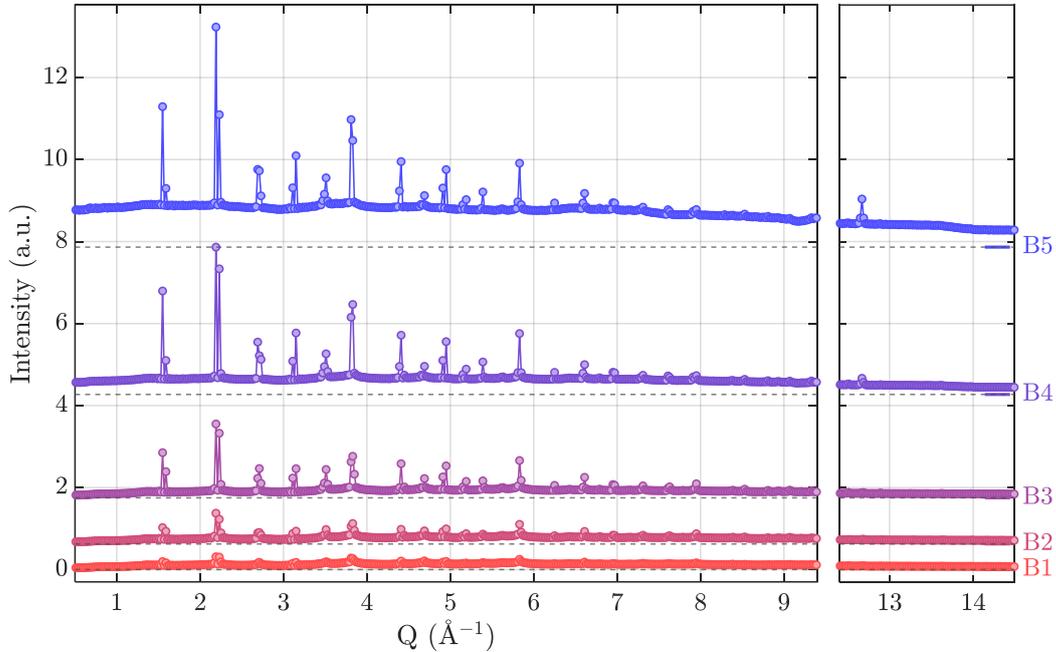

**Figure 5**. $I(Q)$ (Equation 9) for five absorbers (B1, B2, B3, B4, B5 - from bottom to top), offset along the y-axis for clarity. In this series of absorbers, B1 is the weakest and B5 is the strongest. Horizontal dashed lines mark the zero-intensity baseline for each curve. The beam absorber for each intensity trace is marked on the right. This plot illustrates changes in the relative intensities of the Bragg peaks because of saturation effects if using weaker absorbers, while also highlighting artefacts in the diffuse background intensities arising from poor counting statistics for the stronger absorbers.

**Step 2: Introducing the calculated and observed scattering functions**

The original definition of the calculated scattering function, as given by (Farrow & Billinge, 2009) is:

$$S^{\text{calc}}(Q) = \frac{I_{\text{coh}}(Q) - I_{\text{compt}}(Q)}{\langle f \rangle^2 N} - I_{\text{Laue}}(Q) \tag{10}$$

where $I_{\text{coh}}(Q)$ is the spherically averaged coherent scattering intensity, $I_{\text{compt}}(Q)$ represents the inelastic Compton scattering, $I_{\text{Laue}}(Q)$ is the monotonic (Laue) diffuse scattering in a multicomponent system:

$$I_{\text{Laue}}(Q) = \frac{\langle f^2 \rangle}{\langle f \rangle^2} - 1 \tag{11}$$

We define the **observed** scattering function $S^{\text{obs}}(Q)$ as:

$$S^{\text{obs}}(Q) = \frac{s}{\langle f \rangle^2} \left( I_{\text{cryst}}(Q) - p\, I_{\text{compt}}(Q) \right) - I_{\text{Laue}}(Q) \tag{12}$$

where $p$ is the scale factor reflecting the contribution of Compton scattering. The intensity of the latter, $I_{\text{compt}}(Q)$, is calculated according to (Bikondoa & Carbone, 2021)) as described in Appendix C.

**Step 3: Determination of the scale coefficients $s$ and $p$**

The calculated scattering function $S^{\text{calc}}(Q)$ asymptotes to unity at high $Q$. Thus, $s$ and $p$ can be determined by imposing the same requirement on $S^{\text{obs}}(Q)$. Measurements have to extend to $Q$ values sufficiently large for $S(Q)$ to attain this asymptote – a condition that can be verified by considering the expected baseline of the scattering function, $S^{\text{calc}}_{\text{base}}(Q)$:

$$S^{\text{calc}}_{\text{base}}(Q) = 1 - \frac{\langle (fT)^2 \rangle}{\langle f \rangle^2} \tag{13}$$

where

$$\langle (fT)^2 \rangle = \frac{1}{N_u} \sum_{\mu=1}^{N_u} O_\mu (f_\mu T_\mu^s)^2 \text{ and } \langle f \rangle^2 = \left( \frac{1}{N_u} \sum_{\mu=1}^{N_u} O_\mu f \right)^2. \tag{14}$$

Here, $T_\mu^s = T_\mu^s(Q) = \exp\left\{-\frac{1}{2} U_{iso,\mu} Q^2\right\}$ is the orientational average of the generally anisotropic Debye-Waller factor, $T_\mu = T_\mu(\boldsymbol{Q})$ and $U_{iso,\mu}$ is the equivalent isotropic atomic displacement parameter for site $\mu$. $O_\mu$ are the occupancies of these sites. Calculating $S^{\text{calc}}_{\text{base}}(Q)$ requires $U_{iso,\mu}$. The supplementary information for this article includes a script that calculates $S^{\text{calc}}_{\text{base}}(Q)$, given the chemical composition and $U_{ij}$ tensors.

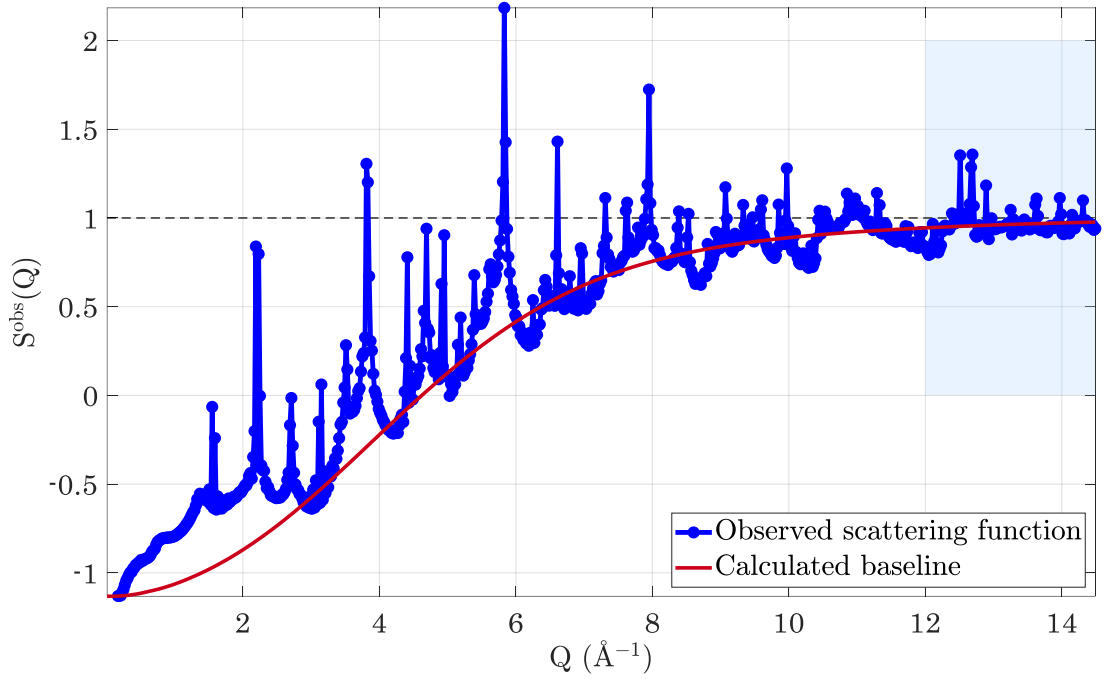

**Figure 6**. Normalized scattering function $S^{obs}(Q)$ as defined in the equation coefficients $s$ and $p$ were calculated such that the resulting $S^{obs}(Q)$ asymptotically approaches 1 at high $Q$. The rectangle on the right highlights the Q-range, which was used for the determination of these coefficients. The red dashed line shows the baseline, calculated according to Equations (13) and (14).

In our case, the conditions were implemented by selecting all data points with $Q > 12$ Å$^{-1}$ (marked by the rectangle in Figure 6), and using linear regression for matching $S^{obs}(Q)$ to $S^{calc}_{base}(Q)$ to find $s$ and $p$. Then, we normalized our reconstructed total intensities per atom according to

$$I(\mathbf{Q}) \leftarrow s(I(\mathbf{Q}) - pI_{compt}(Q)). \tag{15}$$

The relative contribution of the Compton scattering to the total intensity can reach ≈ 50 % at $Q$ = 14 Å$^{-1}$ (Appendix C). Subtracting this parasitic background is essential for the quantitative fitting of diffuse intensities to $Q$ > 5 Å$^{-1}$.

### 3.4 Determining Bragg intensities

The RSR process does not distinguish between the Bragg and diffuse scattering components. However, combining these signals within the same look-up table poses challenges that must be addressed if fitting this table as a single dataset. The difficulty arises because Bragg peaks are typically confined to a single detector pixel and have an angular width that is 1-2 orders of magnitude smaller than Δ$\varphi$. As a result, experimental data provide Bragg-peak intensities integrated over the corresponding voxel volumes.

Below, we describe a procedure for extracting observed Bragg intensities $I^{obs}_{hkl}$ that can be matched to those calculated from a structural model:

$$I_{hkl}^{\text{calc}} = \frac{1}{N_u} \left| \sum_{\mu=1}^{N_u} f_\mu T_\mu O_\mu \exp\left(2\pi i (hx_\mu + ky_\mu + lz_\mu)\right) \right|^2 \quad (16)$$

Here, $N_u$ is the number of atoms in the crystallographic unit cell, $T_\mu$ are the Debye-Waller factors associated with atomic sites having coordinates $x_\mu, y_\mu, z_\mu$. It is possible to show (see e.g. (Warren, 1990)) that

$$I_{hkl}^{\text{calc}} = \iiint_{hkl} I_{\text{coh}}(H, K, L) dH dK dL \quad (17)$$

where the integration is carried out over the Bragg peak $hkl$.

The first step of the proposed procedure involves subtracting the diffuse-scattering background $I_0(H, K, L)$ from the total reconstructed intensity $I_{\text{cryst}}(H, K, L)$. At this stage, the background intensity $I_0(H, K, L)$ beneath the Bragg reflection is estimated by fitting a 3D pseudo-Voigt function to the portion of $I_{\text{cryst}}(H, K, L)$ located at a distance greater than 0.06 r.l.u. from the peak center. Figure 7 illustrates this background subtraction procedure. It represents a 3D reconstruction of $I_{\text{cryst}}(H, K, L)$ around the 011 reflection, including the $HK$, $HL$, $KL$ projections (integrated along the missing coordinate in the range of 0.05 r.l.u.), and $H, K$ and $L$ intensity profiles (integrated over the two missing coordinates). Each 1D plot also includes the corresponding fitted background profile $I_0(H, K, L)$. Then the observed Bragg intensities are calculated as follows:

$$I_{hkl}^{\text{obs}} = \Delta H \, \Delta K \, \Delta L \sum_{HKL \in hkl} (I_{\text{cryst}}(H, K, L) - I_0(H, K, L)) \quad (18)$$

Here, the product $\Delta H \, \Delta K \, \Delta L$ defines the voxel volume, and the summation is performed over those voxels in the reciprocal unit cell centered on the reciprocal lattice node $[hkl]^*$ that that fall within the range ($|H - h|$, $|K - k|$, $|L - l| < 0.5$.

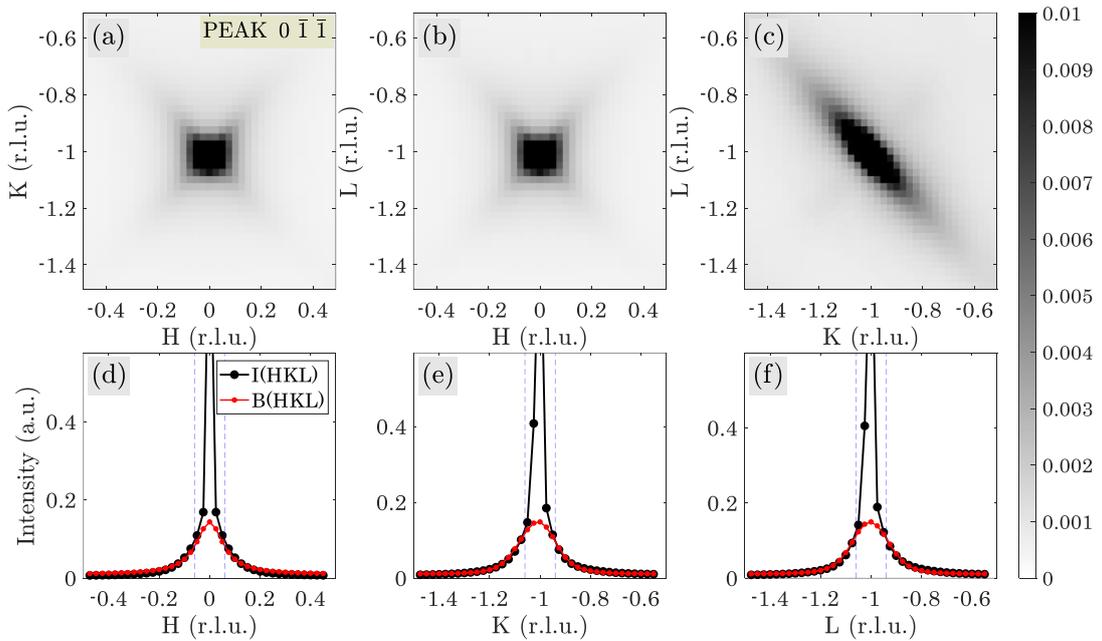

**Figure 7.** The reconstructed scattering intensity around 011 Bragg reflection. The top panels show $HK$, $HL$ and $KL$ projections, while the bottom panels show the $H$-, $K$- and $L$- dependencies, along with the fitted background curve $B_{\text{cryst}}(H, K, L)$.

We use equation (18) to generate tables of $I_{hkl}^{\text{obs}} \pm \sigma_{hkl}^{\text{obs}}$ for each measured reflection and each absorber. We then examine these intensities versus the beam monitor value $B$. Because the RSR normalizes the intensities by $B$ (equations (3)-(4)), unsaturated reflections are flat with respect to B. In contrast, saturated reflections display underestimated intensities at large $B$, where the primary beam is more intense.

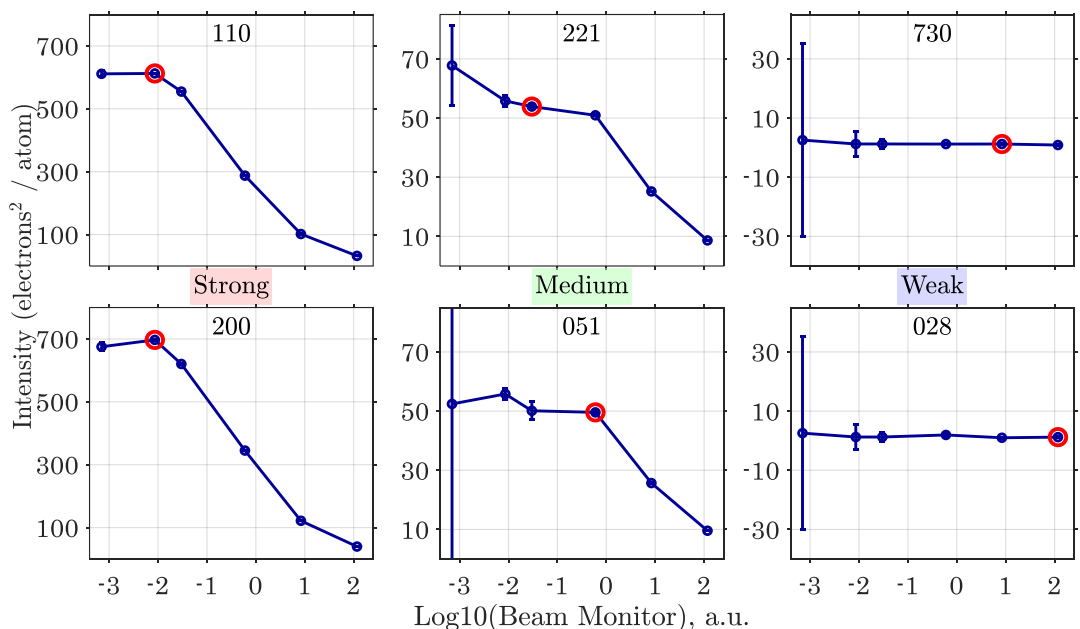

**Figure 8.** Dependence of reflection intensities on the primary beam monitor value $B$, shown on a logarithmic scale, for three groups of representative reflections: strong (110, 200), medium (221, 051), and weak (730, 028). Each panel displays the integrated intensity $I_{hkl}^{\text{obs}}$ with error bars as a function of $\log_{10} B$. For strong reflections, intensities decrease at higher beam monitor values due to detector saturation. Conversely, weak reflections exhibit low signal-to-noise ratios at low flux. Red circles indicate the absorber setting automatically selected for each reflection by the procedure used to optimize data quality for structural refinement.

Figure 8 plots the examples of $I_{hkl}^{\text{obs}}$ versus B for three reflection groups: strong (110 and 200), medium (221 and 051), and weak (730 and 028). As expected, strong peaks exhibit reduced intensities at large B due to detector saturation; for these reflections, we select intensities from datasets recorded with a higher-attenuation absorber. Conversely, weak reflections require higher-flux data to achieve adequate SNR. We automated this per-reflection choice by inspecting $I_{hkl}^{\text{obs}} \pm \sigma_{hkl}^{\text{obs}}$ as a function of B; in Figure 8, red circles mark the absorber selected for each reflection. The intensity values for those voxels in the total scattering dataset that contain Bragg peaks can now be replaced with the corresponding unsaturated values.

### 4. Crystallographic refinements

All refinements were performed in ShelXle (Hübschle *et al.*, 2011; Sheldrick, 2015). We considered two Bragg intensity datasets: one from the laboratory instrument with a proven track record of providing reliable structural models and another from our synchrotron measurements with the intensities extracted as described in Section 3.4. The initial model assumed a perovskite $Pm\bar{3}m$ space group with all the atoms residing at the ideal positions: Pb @ 1*a* 000; Mg and Nb@1*b* ½½½; O @ 3c ½½0. This model produced a poor fit and anomalously large atomic displacement parameter for Pb for both datasets (laboratory data: *wR2* ≈ 16 %, $U_{Pb}$ ≈ 0.08 Å$^2$; synchrotron data: *wR2* ≈ 20 %, $U_{Pb}$ ≈ 0.08 Å$^2$).

We then considered two models with Pb disordered over multiple sites offset from the ideal central position. One assumed Pb atoms shifted along the ⟨100⟩ directions (Wyckoff position 6*e*: $x$ 0 0) and another – along the ⟨111⟩ directions (Wyckoff position 8*g*: $x\ x\ x$). Both split-site models produced a significant improvement in the fit with similar agreement factors for each dataset. For the laboratory data, *wR2* ≈ 7.2 % and *wR2* ≈ 6.98 % for the 8- and 6-site Pb models, respectively. Regardless of the displacement directions, the magnitude of the Pb off-centering relative to the centrosymmetric positions was ≈ 0.28 Å. For the synchrotron data, the agreement factors were considerably worse, with wR2 ≈ 9.2 % for both 8-site and 6-site models. Refining the Mg/Nb ratio using the laboratory data yielded 0.54(2), which is reasonably close to the expected value of 0.5. For the synchrotron data, the refined ratio significantly deviated from the expected value (i.e., ≈ 1 instead of 0.5); therefore, in the final refinement, we kept the composition fixed at the stoichiometric value. The 6- and 8-site models were indistinguishable, in line with (Eremenko et al, 2025), who observed the coexistence of both types of Pb displacements.

**Table 1**: Comparison of structural refinements for the 8-site and 6-site Pb models in PbMg$_{1/3}$Nb$_{2/3}$O$_3$ using data from the Bruker XRD and ID28 beamline at ESRF. Space group $Pm\bar{3}m$; lattice parameter $a = 4.0451(1)$ Å. In the 8-site model, Pb atoms occupy $x\ x\ x$ positions; in the 6-site model, they occupy $x\ 0\ 0$ positions. Atomic displacement parameters $U_{ij}$ are given in Å$^2$; for Nb and Mg, $U$ is isotropic and constrained to be equal for both species. Values in parentheses represent one standard deviation in the last significant digit, as estimated by ShelXle. For the ID28 refinement, the Nb site fraction was fixed.

| Parameter | $x_{Pb}$ | Nb frac | $U_{11}(Pb)$ | $U_{23}(Pb)$ | $U(Nb)$ | $U_{11}(O)$ | $U_{33}(O)$ |
|---|---|---|---|---|---|---|---|
| 8-site | | | | | | | |
| Bruker | 0.0391(3) | 0.652(14) | 0.0228(9) | -0.0059(3) | 0.0085(5) | 0.011(2) | 0.026(2) |
| ID28 | 0.0391(4) | 0.667 | 0.0237(8) | -0.0064(6) | 0.0102(5) | 0.011(2) | 0.028(2) |
| 6-site | | | | | | | |
| Parameter | $x_{Pb}$ | Nb frac | $U_{11}(Pb)$ | $U_{33}(Pb)$ | $U(Nb)$ | $U_{11}(O)$ | $U_{33}(O)$ |
| Bruker | 0.0684(4) | 0.652(14) | 0.011(1) | 0.027(1) | 0.0085(5) | 0.010(2) | 0.027(2) |
| ID28 | 0.0694(8) | 0.667 | 0.013(1) | 0.030(1) | 0.0114(7) | 0.008 (3) | 0.030(4) |

Despite the significantly poorer agreement factors for the synchrotron data, the structural parameters, including the magnitude of the Pb off-centering and the $U_{ij}$ values for Mg/Nb and O sites were similar to those obtained from the laboratory data (Table 1), indicating that the structural parameters, in our case, are robust, even in the presence of apparently larger systematic errors. The existence of larger errors in the synchrotron data was also manifested in the relatively large extinction coefficient (1.34) compared to a much more reasonable value of 0.35 obtained while fitting the laboratory dataset.

5. Scale factor

Given that the synchrotron data were normalized to an absolute scale per atom, the expected scale factor is $s$ = 0.2 (5 atoms per unit cell), whereas our refinements return $s \approx$ 0.35. Despite the ≈1.5-fold difference, the agreement is deemed satisfactory considering the measurement limitations and the complexity of data processing. This result validates the scaling procedure based on the asymptotic behavior of the scattering function.

In ShelXle refinements, the scale factor is strongly correlated with the extinction coefficient. In our case, for the synchrotron data, the correlation coefficient between these two parameters was $\approx$ 0.82. For the laboratory data, the correlation was even stronger with this coefficient $\approx$ 0.92. Hence, having a separate scaling procedure based on the predicted baseline of the scattering function provides a check on the scale factor, improving the reliability of scaled intensities, both Bragg and diffuse. While using Bragg and diffuse components obtained in the same measurements carries obvious advantages, especially if performing in situ measurements, the availability of the validated scaling procedure just for the diffuse part permits combining Bragg and diffuse datasets measured using different instruments.

**Summary**

We developed a procedure for obtaining Bragg and diffuse X-ray scattering intensities in the same 3D dataset. Experimental measurements were performed using a synchrotron beamline optimized for recording weak diffuse scattering. The currently inevitable detector saturation by strong Bragg reflections was addressed by collecting data while using a series of absorbers covering five decades in intensity. We developed robust protocols for identifying voxels in the reconstructed 3D intensity distributions that were affected by saturation and replacing their intensities with those from datasets collected using the appropriate absorber.

We then compared two scaling methods for relating measured intensities to their absolute-scale (electrons$^2$/atom) theoretical values. The first method uses traditional crystallographic refinements, matching Bragg intensities calculated for the average structure model to the observed values. The second, introduced recently but still unverified, matches the 1D scattering function calculated as a spherical average of measured scattered intensities to a theoretical baseline estimated from the chemical composition and atomic displacement parameters. The latter function has a known behavior at large momentum transfers that can be used to scale experimental data; this scaling can be performed on diffuse scattering alone, without including Bragg reflections. Our results verified that both methods yield similar scale factors, providing a firm ground for the use of the calculated scattering-function baseline to scale the diffuse portion of the total signal.

With the ability to scale Bragg peaks and diffuse scattering independently, it becomes possible to combine these datasets in structural refinements even if they have been measured using different instruments. For example, a viable approach would be to collect Bragg data using a laboratory diffractometer and diffuse scattering data at a synchrotron or in an electron microscope. A capability for such refinements using a list of observed structure factors and 3D diffuse scattering as input has been implemented in RMCProfile, and their examples will be reported separately.

**Acknowledgements**

The crystals used in this study were provided by Z-G Ye (Simon Fraser University). We thank the European Synchrotron Radiation Facility for providing beam time on ID28 and D. Chernyshov for providing advice on data collection and structural refinements. The research by S. G. and I. L. was supported by the US-Israel Binational Science Foundation (Award No. 2018161). S.G. acknowledges the support of Israel Science Foundation (Grant Nos. 1561/18, 3455/21, 1365/23)

**Appendix A: Conversion between the instrumental and scattering vector coordinates**

This appendix summarizes the relationship between the components of the scattering vector $\boldsymbol{Q} = (Q_x, Q_y, Q_z)$ and the instrumental coordinates $x_d, y_d, \varphi, \omega, \chi$ where:

- $x_d, y_d$ are detector-plane coordinates of a given detector pixel,
- $\varphi$ is the crystal rotation angles around the corresponding Eulerian cradle axis,
- $\omega$ and $\chi$ are the other two Eulerian angles (typically held constant during the acquisition).

The scattering vector $\boldsymbol{Q}$ is defined in the laboratory coordinate system corresponding to the diffractometer configuration where all Eulerian angles $(\omega, \chi, \varphi)$ are set to zero.

We use the standard definition of the scattering vector:

$$\boldsymbol{Q}' = \frac{2\pi}{\lambda}(\boldsymbol{e}_1 + \boldsymbol{d}) \qquad (19)$$

Here

- $\lambda$ is the X-ray wavelength,
- $\boldsymbol{e}_1$ is the unit vector pointing from the crystal to the X-ray source
- $\boldsymbol{d}$ are the unit vector pointing from the crystal to the detector pixel.

The vector $\boldsymbol{d} = \frac{\boldsymbol{D}_{xy}}{D_{xy}}$ is given by the normalized detector-pixel position vector $\boldsymbol{D}_{xy}$:

$$\boldsymbol{D}_{xy} = D_0 \boldsymbol{d}_0 + x\boldsymbol{x}_d + y\boldsymbol{y}_d \qquad (20)$$

- $D_0$ is the distance from the sample to the detector plane
- $\boldsymbol{d}_0$ is the unit vector which is normal to the detector plane

- $\mathbf{x}_d$ and $\mathbf{y}_d$ are the unit vectors defining the detector axes.

Equations (19) and (20) allow to calculate the coordinates of the vector $\mathbf{Q}' = (Q'_x, Q'_y, Q'_z)$. To obtain the corresponding scattering vector in the unrotated laboratory frame, we apply the inverse of the rotation matrix $[M_{\omega\chi}(\varphi)]$

$$\begin{pmatrix} Q_x \\ Q_y \\ Q_z \end{pmatrix} = [M_{\omega\chi}(\varphi)]^{-1} \begin{pmatrix} Q'_x \\ Q'_y \\ Q'_z \end{pmatrix} \quad (21)$$

The total rotation matrix $[M_{\omega\chi}(\varphi)]$, as defined in e.g. (Gorfman *et al.*, 2021), is the product of three elementary rotations.

$$[M_{\omega\chi}(\varphi)] = \begin{pmatrix} \cos\omega & \sin\omega & 0 \\ \sin\overline{\omega} & \cos\omega & 0 \\ 0 & 0 & 1 \end{pmatrix} \begin{pmatrix} 1 & 0 & 0 \\ 0 & \cos\chi & \sin\chi \\ 0 & \sin\overline{\chi} & \cos\chi \end{pmatrix} \begin{pmatrix} \cos\varphi & \sin\varphi & 0 \\ \sin\overline{\varphi} & \cos\varphi & 0 \\ 0 & 0 & 1 \end{pmatrix} \quad (22)$$

All the coordinate axes are summarized in Figure 1.

**Appendix B: Relationship Between Measured and Theoretical intensities**

This appendix summarizes key elements of the kinematical X-ray scattering theory that relate the measured intensity, $J_{\text{lab}}(x_d, y_d, \varphi)$ to the calculated coherent scattering intensity $I_{\text{coh}}(\mathbf{Q})$.

We begin by defining the measured quantity $J_{\text{lab}}(x_d, y_d, \varphi)$, which represents the number of photons accumulated at the detector pixel $(x_d, y_d)$ during the continuous rotation of the crystal over an angular range $[\varphi, \varphi + \Delta\varphi]$ and exposure time $\tau$:

$$J_{\text{lab}}(x_d, y_d, \varphi) = \frac{\tau}{\Delta\varphi} \int_{\varphi}^{\varphi+\Delta\varphi} J_{\text{flux}}(x_d, y_d, \varphi) \, d\varphi \quad (23)$$

Here $J_{\text{flux}}(x_d, y_d, \varphi)$ is the instantaneous photon flux (in photons / s) arriving at the pixel during the rotation. This flux can be related to the local flux density $J_{\text{fd}}(x, y, \varphi)$ (in photons / s / mm²) via integration over the pixel area $S$, accounting for the oblique incidence angle $\eta$ (the angle between the scattered ray and the detector normal):

$$J_{\text{flux}}(x_d, y_d, \varphi) = \cos\eta \iint_{\text{PIXEL}} J_{\text{fd}}(x, y, \varphi) dx dy \quad (24)$$

According to the kinematical theory of X-ray scattering (Warren, 1990; Guinier, 1994; Als-Nielsen & McMorrow, 2011), the local flux density is given by

$$J_{\text{fd}}(x, y, \varphi) = J_0 \frac{r_e^2}{D^2} P(x, y) |A'(x, y, \varphi)|^2 \quad (25)$$

Here

- $J_0$ is the incidence beam flux density (measured in photons / s / mm²),
- $r_e$ is the classical electron radius,
- $D$ is the sample-to-pixel distance,
- $P(x, y) = \sin^2\psi(x_d, y_d)$ is the polarization factor.
- $A'(x, y, \varphi)$ is the scattering amplitude corresponding to detector position $(x, y)$ at rotating angle $\varphi$.

- $\psi(x_d, y_d)$ is the angle between the polarization direction and the scattered beam.

To simplify notation, we define the sample-to-detector distance as $D_0 = D \cos \eta$ and rewrite (23) as:

$$J_{\text{lab}}(x_d, y_d, \varphi) = J_0 \frac{r_e^2}{D_0^2} \frac{\tau}{\Delta \varphi} \cos^3 \eta \, P \iiint |A'(x, y, \varphi)|^2 dx dy d\varphi \qquad (26)$$

Next, we use the functional dependence between the instrumental $(x, y, \varphi)$ and the scattering vector $(Q_x, Q_y, Q_z)$ in the unrotated reference frame as described in Appendix A as $\mathbf{Q}(x, y, \varphi)$. Substituting $A'(x, y, \varphi) = A(Q_x, Q_y, Q_z)$, equation (26) becomes

$$J_{\text{lab}}(x_d, y_d, \varphi) = J_0 \frac{r_e^2}{D_0^2} \frac{\tau}{\Delta \varphi} \cos^3 \eta \, P \iiint |A(Q_x, Q_y, Q_z)|^2 dx dy d\varphi \qquad (27)$$

We now introduce the generalized Lorentz as the inverse Jacobian determinant:

$$L^{-1}(x, y, \varphi) = \left| \frac{\partial(Q_x, Q_y, Q_z)}{\partial(x, y, \varphi)} \right| \qquad (28)$$

Which allows a change of variables:

$$dx dy d\varphi = L(x, y, \varphi) dQ_x dQ_y dQ_z,$$

Substituting this into Eq. (27), we obtain

$$J_{\text{lab}}(x, y, \varphi) = J_0 \frac{r_e^2}{D_0^2} \frac{\tau}{\Delta \varphi} \cos^3 \eta \, \text{LP} \iiint |A(Q_x, Q_y, Q_z)|^2 dQ_x dQ_y dQ_z \qquad (29)$$

Although, both the Lorentz factor L and polarization factor P vary across the detector, they can be treated as constant over the small reciprocal-space volume $\delta V^*$ covered by the single 3D pixel. This allows an alternative expression for the Lorentz factor in terms of the volume ratios:

$$L(x, y, \varphi) = \frac{\Delta \varphi \, S}{\delta V^*} \qquad (30)$$

So that equation (27) is transformed into:

$$J_{\text{lab}}(x_d, y_d, \varphi) = J_0 \frac{r_e^2}{D_0^2} \frac{S\tau}{\delta V^*} \cos^3 \eta \, P \iiint |A(Q_x, Q_y, Q_z)|^2 dQ_x dQ_y dQ_z \qquad (31)$$

Applying the multiplicative correction factor as defined in (3) and (4) of the main text:

$$R(x_d, y_d, \varphi) = \frac{R_0 D_0^2}{B \tau \sin^2 \psi \cos^3 \eta} \, \delta V^*$$

We obtain the corrected scattering intensity as

$$J'_{\text{lab}}(x_d, y_d, \varphi) = \iiint |A(Q_x, Q_y, Q_z)|^2 dQ_x dQ_y dQ_z \qquad (32)$$

The reconstruction procedure was introduced in the main text: it involves converting the corrected intensity $J'_{\text{lab}}(x_d, y_d, \varphi)$ into $I_{\text{cryst}}(H, K, L)$

$$I_{\text{cryst}}(H,K,L) = \frac{\sum J'_{\text{lab}}(x_d, y_d, \varphi)}{\sum \delta V^*}$$

Which means

$$I_{\text{cryst}}(H,K,L) = R_0^{-1} \langle |A(\boldsymbol{Q})|^2 \rangle \tag{33}$$

where the averaging is performed over the reciprocal-space voxel corresponding to the coordinates $HKL$.

Finally, we recall the standard kinematical theory expression for the scattering amplitude

$$A(\boldsymbol{Q}) = \sum_m f_m(Q) \exp(i\boldsymbol{Q}\boldsymbol{R}_m), \tag{34}$$

which is directly connected to the definition of the coherent scattering intensity (2):

$$I_{\text{coh}}(\boldsymbol{Q}) = \frac{1}{N}|A(\boldsymbol{Q})|^2 \tag{35}$$

Which leads to Eq (7) in the main text:

$$I_{\text{cryst}}(H,K,L) = s \langle I_{\text{coh}}(H,K,L) \rangle, \qquad s = NR_0^{-1} \tag{36}$$

**Appendix C. Compton scattering**

The intensity of Compton scattering is calculated as

$$I_{\text{compt}}(Q) = K(Q) S_{\text{inc}}(Q) \tag{37}$$

Where $K(Q)$ is known as the Klein-Nishina factor

$$K(Q) = \frac{E'}{E}\left(\frac{E}{E'} + \frac{E'}{E} - 2\sin^2 2\theta \cos^2 \psi\right) \tag{38}$$

Here $E$ and $E'$ are the energies of the incident and incoherently scattered photons, respectively. $E'$ is determined through:

$$\frac{E}{E'} = 1 + \frac{E}{E_e}(1 - \cos 2\theta) \tag{39}$$

and $E_e = 0.511$ MeV is the rest energy of a free electron.

The incoherent scattering function $S_{\text{inc}}(Q)$ is evaluated as an incoherent sum of contributions from all the atoms in the unit cell

$$S_{\text{inc}}(Q) = \sum_{\mu=1}^{N_u} O_\mu \, s_{\text{inc},\mu}(Q)$$

where the sum runs over $N_u$ atoms in the unit cell, $O_\mu$ denotes the partial occupancy of the $\mu$-th atom, and $s_{\text{inc},\mu}(Q)$ is the tabulated incoherent scattering factor for the corresponding chemical element (Balyuzi, 1975).

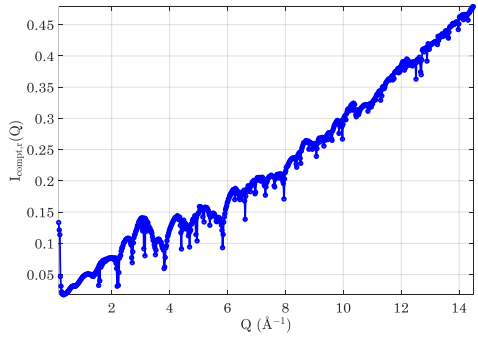

**Figure 9**. $Q$-dependence of Compton scattering contribution defined as $I_{compt,r}(Q) = \frac{pI_{\text{compt}}(Q)}{I(Q)}$. The graph shows that at high $Q$ values, Compton scattering can contribute significantly — up to 50% to the total scattering.